\documentclass[aps,prb,twocolumn,amssymb,color,pdflatex]{revtex4}

\usepackage{graphicx}
\usepackage{graphics}
\usepackage{dcolumn}
\usepackage{bm}
\usepackage{amsmath}
\begin{document}

\newcommand{\bk}{{\bf k}}
\newcommand{\bp}{{\bf p}}
\newcommand{\bv}{{\bf v}}
\newcommand{\bq}{{\bf q}}
\newcommand{\tbq}{\tilde{\bf q}}
\newcommand{\tq}{\tilde{q}}
\newcommand{\bQ}{{\bf Q}}
\newcommand{\br}{{\bf r}}
\newcommand{\bR}{{\bf R}}
\newcommand{\bB}{{\bf B}}
\newcommand{\bA}{{\bf A}}
\newcommand{\bE}{{\bf E}}
\newcommand{\bj}{{\bf j}}
\newcommand{\bK}{{\bf K}}
\newcommand{\cS}{{\cal S}}
\newcommand{\vd}{{v_\Delta}}
\newcommand{\tr}{{\rm Tr}}
\newcommand{\kslash}{\not\!k}
\newcommand{\qslash}{\not\!q}
\newcommand{\pslash}{\not\!p}
\newcommand{\rslash}{\not\!r}
\newcommand{\bs}{{\bar\sigma}}
\newcommand{\omt}{\tilde{\omega}}

\title{Vortex as a probe - suggested measurement of the order parameter structure in iron-based superconductors}

\author{Eugeniu Plamadeala, T. Pereg-Barnea, and Gil Refael}
\affiliation{Department of Physics,
California Institute of Technology, 1200 E. California Boulevard, MC114-36,
Pasadena, California 91125, USA }

\date{\today}

\begin{abstract}
Impurities, inevitably present in all samples, induce elastic transitions between quasiparticle states on the contours of constant energy.  These transitions may be seen in Fourier transformed scanning tunneling spectroscopy experiments, sorted by their momentum transfer. In a superconductor, anomalous scattering in the pairing channel may be introduced by magnetic field.  When a magnetic field is applied, vortices act as additional sources of scattering.  These additional transition may enhance or suppress the impurity induced scattering.  We find that the vortex contribution to the transitions is sensitive to the momentum-space structure of the pairing function.  In the iron-based superconductors there are both electron and hole pockets at different regions of the Brillouin zone.  Scattering processes therefore represent intra- or inter-pocket transitions, depending on the momentum transfer in the process. In this work we show that while in a simple $s$-wave superconductors all transitions are enhanced by vortex scattering, in an $s_\pm$ superconductor only intra-pocket transitions are affected.  We suggest this effect as a probe for the existence of the sign change of the order parameter.
\end{abstract}
\maketitle
\section{Introduction}
In January 2008, Kamihara {\it et al.}\cite{DiscoveryFepnictides} announced the discovery of superconductivity in La[O$_{1-x}$F$_x$]FeAs and shortly after more compounds of the iron-based superconductors (FeSC) family were discovered.  This family shares a number of important characteristics with the high T$_c$ cuprates such as the layered structure and the proximity of the superconducting phase to a magnetic one.  Given these similarities it is natural to ask whether the FeSC are {\it conventional} or {\it unconventional}.  By "conventional" it is usually meant that the pairing mechanism is based on the interaction of fermions and phonons with a rotationally symmetric order parameter (OP), as described by the BCS theory\cite{bcs}.  An "unconventional" superconductor may result from any other mechanism and its OP would have a non-trivial structure.  This question is not easily answered since one can not probe the underlying state, i.e., the state of the system without the pairing instability.  A related question seems to be an easier starting point; what is the structure of the pairing function? In particular we would like to be able to distinguish between a simple $s$-wave order parameter and other, more complex OPs.  A simple $s$-wave would very likely deem the FeSC family conventional and may restrict T$_c$.  Any other result will suggest an unconventional pairing mechanism involving electric, magnetic and/or lattice interactions.

The purpose of this paper is to propose an experiment to distinguish between two prominent candidates for the order parameter in the iron based superconductors.  As we discuss below, this may prove to be a difficult task.
Armed with the experience of identifying the $d$-wave OP in the cuprates\cite{PhaseExperiment,HardyBonn-dwave} and state of the art probes we are in a good position to distinguish between a simple $s$-wave structure and higher angular momentum OPs.  The distinction is usually made through the observation (or lack) of nodal quasiparticles - low energy excitations that reside in the vicinity of the intersection between the Fermi surface and the OP nodal lines.  Such nodal quasiparticles may be seen in thermodynamic properties such as transport or NMR relaxation rate.  In the FeSCs, early NMR/NQR experiments have reported the absence of coherence peaks and a power law behavior of $1/T_1 T$ as a function of temperature,  which have been formerly related to the nodal quasiparticles\cite{Nakai,Grafe,Matano,Mukuda,Fletcher}.  However, it seems that these facts alone do not necessarily imply gapless excitations when more than one band is involved\cite{Kawasaki}.  Other experiments such as ARPES\cite{Ding,Takeshi,Zhao}, microwave penetration depth\cite{Hashimoto,Malone} and others report nodeless gaps.  To date, it seems that apart from LaFePO which has a $d$-wave order parameter\cite{MolerLaFePO} other compounds have nodeless OPs.

Based on the above findings we choose to focus on OPs with $s$-wave symmetry (in the lattice this would be a discrete rotational symmetry). The FeSC has five electrons in the $d$ shell of the iron and its low energy band structure is composed of two concentric hole pockets around the Brillouin zone center (the $\Gamma$ point) and two electron pockets around $(0,\pi)$ and $(\pi,0)$ (the $M$ points)\cite{TwoBandModel}.  It is therefore possible that the order parameter changes its magnitude and phase between these two bands.  In this work, we propose an experiment to distinguish between a simple $s$-wave, which does not change sign on or between the bands, and the so called $s_\pm$ OP.  The latter order parameter can be roughly sketched as a single function of momentum, $\Delta_\pm(\bk) = \Delta_0\cos(k_x)\cos(k_y)$\cite{Mazin} which has line nodes {\em between} the electron and hole pockets.  It therefore gaps both Fermi surfaces and does not allow for nodal excitations.  However, the sign of the order parameter changes from one pocket to the other.  We focus on this OP since it arises from several microscopic models such as extended t-J model\cite{tJModel}, FS nesting and exchange interactions\cite{Mazin,Arita}, an interaction induced density wave \cite{CvetkovicTesanovic} 
and functional renormalization group of a strongly interacting lattice model\cite{DungHai}.

The $s_\pm$ order parameter poses a challenge for the experimental probes.  Since both Fermi pockets are fully gapped, experiments which are mainly sensitive to the spectrum (ARPES, NMR etc.) are incapable of detecting its sign change.  On the other hand, a phase sensitive probe such as the one devised by Tsuei and Kirtley  for the cuprates\cite{PhaseExperiment} is not easily achieved since the OP sign depends on the {\em amplitude} of the momentum rather than its direction\cite{Parker,Wu}.  A natural route to explore is then the effect of intrinsic or induced scattering processes.  We show below how intra- and inter-pocket scattering processes may be identified in Fourier transformed (FT) scanning tunneling spectroscopy experiments.  The scattering probability is sensitive to both the sign and magnitude of the OP.  The response of these transitions to vortex scattering may reveal the sign difference between the initial and final states and allow the distinction between a simple $s$-wave and an $s_\pm$ OP.

This article is organized as follows.  In the next section we sketch the suggested experiment and its interpretations; in section III we describe our minimal two-band model framework\cite{TwoBandModel} and calculate the local density of states (LDOS) modulations which arise from both impurity and vortex scattering\cite{TPBMFfield}; in section IV we present and discuss our results.
\section{Proposed experiment}
Spatial modulations in the LDOS are a signature of disorder.  We model the disorder as a point-like potential (impurity or vortex) and perform the Born approximation. When the LDOS modulations are measured at energy $\hbar\omega$ the relevant processes (elastically) take a quasiparticle from a momentum state $\bk_i$ to a momentum state $\bk_f$.  The largest contribution to the LDOS modulations comes from the vicinity of the relevant contours of constant energy.  When the LDOS is Fourier transformed (FT) a feature appears at any momentum $\bq = \bk_f-\bk_i$ which matches two points on the contours of constant energy\cite{Hoffman}.  Each process's contribution is further weighted by quantum mechanical considerations such as the phase space available for scattering\cite{egGraphene} and the scattering potential matrix elements between the initial and final wave functions.  In superconductors, these matrix elements depend crucially on the amount of particle-hole mixing in the state, i.e., the Bogoliubov-de Gennes coherence factors\cite{TPBMFrapid} which, in turn, depend on the magnitude and sign of the order parameter.  As suggested by Zhang {\em et al.}\cite{ZhangQPI}, one can recognize the fingerprint of the suggested $s_{\pm}$ OP in the quasiparticle interference maps.  However, we suspect that the signature of the $s_\pm$ state might not be easy to identify in a disordered system.  Our current suggestions builds on the quasiparticle interference maps idea and adds the magnetic field as a knob which will alter features in a way that can reveal the elusive $s_\pm$ state.

\begin{figure}
\includegraphics[width=0.8\columnwidth]{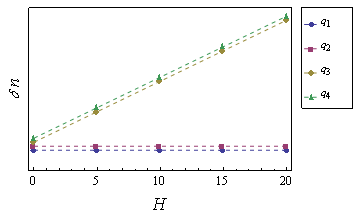}
\caption{{\bf Schematic plot of feature intensity vs. magnetic field } (Color online) The intensity of the Fourier transformed LDOS at momenta $\bq_i$ (defined as momentum transfer in the scattering processes described in Fig.~\ref{fig:CCElabeled}) as a function of magnetic field.  When the magnetic field is applied the vortex induced scattering is added to the impurity scattering.  For intra-pocket transitions ($\bq_3,\bq_4$) the intensity is enhanced and for inter-pocket transitions ($\bq_1,\bq_2$) it is not.}
\label{fig:schematic}
\end{figure}

Our findings indicate that the relative intensity of the inter-pocket and intra-pocket transitions will change dramatically when the magnetic field is turned on if the order parameter changes sign between pockets.  The magnetic field creates vortices which are pinned to the impurity sites.  These vortices act as scattering sources in the particle-particle (off diagonal) channel\cite{TPBMFfield}.  The contribution of the vortices to the LDOS modulations has features in the same momenta $\bq$ as the impurity scattering (since $\bq$ is determined by the contours of constant energy).  The intensity, however, has a different dependence on the coherence factors\cite{TPBMFIJMP}.  In fact, when the energy is tuned to the gap edge, the vortex scatter contribution,$\delta N_v$, is roughly:
\begin{equation}
\delta N_v(\bq,\omega\sim\Delta_0) \propto \omega(\Delta(\bk_i)+\Delta(\bk_f))^2.
\end{equation}
This means that while intra-pocket transitions are always enhanced, inter-pocket transitions are enhanced only if the sign of the OP is the same on both pockets.  In the case of the $s_\pm$ OP $\Delta(\bk_i) \approx -\Delta(\bk_f)$ for inter-pocket transitions which means no vortex contribution to those transitions.  In other words - if the FeSC have an order parameter that changes sign between the electron and hole pockets then the application of magnetic field will affect the intensity of the Fourier transformed LDOS at momenta $\bq_{intra}$ which connect two points on the same pocket and will only weakly affect the intensity of features whose momentum $\bq_{inter}$ connects two points on different pockets.  The qualitative results of the suggested experiments are schematically plotted in Fig.~\ref{fig:schematic} where the feature intensity is modified (or unchanged) when magnetic field is applied due to the addition of vortex scattering.

In the next section we demonstrate this principle through a phenomenological, BCS-like model, based on a two-orbital band structure\cite{TwoBandModel} and the sign-changing $s_\pm$ OP.

\section{Analysis}
\begin{figure}
\includegraphics[width=0.8\columnwidth]{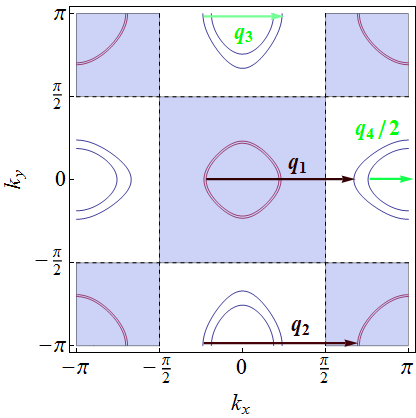}
\caption{{\bf Contours of constant energy. }
(Color online) The iron Brillouin zone with the contours of constant energy, $\omega = 0.105$, right at the gap edge.  The (red) contours around the $\Gamma$ points are the hole pockets and the (blue) contours around the M points are the electron pockets. The vectors $\bq_1$ and $\bq_2$ are inter-pocket transitions while $\bq_3$ and $\bq_4$ are intra-pocket transitions in the hole and electron pockets respectively.  The nodal lines of the $s_\pm$ order parameter are marked by the horizontal and vertical dashed lines and the areas with a positive(negative) OP are shaded(unshaded).
}
\label{fig:CCElabeled}
\end{figure}

In this section we review the two-band model for the FeSC\cite{TwoBandModel} and adopt it as the unperturbed Hamiltonian to describe the uniform system.  We then consider quasiparticle scattering off impurities and vortices using the Born approximation.

\subsection{Two band Model}
The unperturbed Hamiltonian we use is given by:
\begin{eqnarray}
{\cal H}_0^{MF} &=& \sum_k \psi(\bk)^\dagger \hat{h}(\bk) \psi(\bk)   \\
\hat{h}(\bk) &=& \hat{h}_t(\bk) + \Delta(\bk) \sigma_0 \otimes \tau_1 \nonumber
\\ \nonumber
\hat{h}_t(\bk) &=& \left[ \left( \epsilon_+(\bk) - \mu \right)\sigma_0 + \epsilon_-(\bk) \sigma_3 + \epsilon_{xy}(\bk)\sigma_1 \right] \otimes \tau_3
\end{eqnarray}
Where $\psi(\bk)^\dagger = \left( c^\dagger_{\bk,\uparrow}, c_{-\bk, \downarrow}, d^\dagger_{\bk, \uparrow}, d_{-\bk,\downarrow} \right) $ is a vector representing both the orbital and Nambu degrees of freedom. Here $c^\dagger_{\bk, \sigma}$ creates an electron carrying momentum $\bk$ and spin $\sigma$ in the '$d_{xz}$' orbital and $d^\dagger_{\bk, \sigma}$ creates and electron in the '$d_{yz}$' orbital. The $\sigma$ Pauli matrices act in the orbital space and the $\tau$ Pauli matrices act in Nambu space. In what follows we consider the $s_\pm$ OP, $\Delta(\bk) = \Delta_0 \cos k_x \cos k_y$ unless otherwise stated.  The band structure $\hat{h}_t$ is the result of hopping terms of the two orbits on nearest- and next-nearest bonds with the appropriate overlap amplitudes:
\begin{eqnarray}
\epsilon_+(\bk) &=& -(t_1 + t_2) ( \cos k_x + \cos k_y) - 4 t_3 \cos k_x \cos k_y \nonumber
\\
\epsilon_-(\bk) &=& -(t_1 - t_2) (\cos k_x - \cos k_y) \nonumber
\\
\epsilon_{xy}(\bk) &=& -4 t_4 \sin k_x \sin k_y.
\end{eqnarray}
A realistic set of parameters is: $t_1 = -1, t_2 = 1.3, t_3 = t_4 = -0.85, \Delta_0 = 0.1 $ and $\mu$ between $1.3$ and $1.9$, where energy is measured in units of $|t_1|$\cite{TwoBandModel}.

The unperturbed, retarded Green's function $G^0(\bk, \omega) = \left[ (\omega + i \eta) \mathbb{I}_4 - \hat{h}(\bk) \right]^{-1}$.
\subsection{Impurity and vortex scattering potential}
Next we consider a perturbation and its effect on the local density of states.  The perturbing Hamiltonian is of the form:
\begin{equation}\label{eq:perturbation}
\delta{\cal H} = \sum_{\bk,\bk'} d\bk d\bk' \Psi^\dagger_{\bk} V_{\bk,\bk'}\Psi_{\bk'}
\end{equation}
where $V_{\bk,\bk'}$ is a $4\times4$ matrix in the two band Nambu space.  Its matrix  elements are general enough to describe on-site or hopping-like potential in both the particle-hole channel (charge/spin impurity) or the particle-particle channel (pairing perturbation).

In the Born approximation, the perturbed Green's function is given by:
 \begin{eqnarray}\label{eq:GVG}
 G(\bk,\bk',\omega) &=& G^0(\bk,\omega)\delta_{\bk,\bk'} + \delta G(\bk,\bk'\omega) \nonumber \\
\delta G(\bk, \bk', \omega) &=& G^0(\bk, \omega) V_{\bk, \bk'} G^0(\bk', \omega)
\end{eqnarray}

As a result, the induced FT LDOS modulations are given by:\cite{ZhangQPI}
\begin{eqnarray}\label{eq:deltan}
\delta n(\bq, \omega) &=& -\frac{1}{\pi}{\rm Im}\int \frac{d^2 k}{ (2\pi)^2 }\\ \nonumber
 &&\left[\delta G(\bk, \bk+\bq,\omega)_{11} + \delta G(\bk, \bk+\bq,\omega)_{33} \right]
\end{eqnarray}
We classify the various potentials in Eq.~\ref{eq:GVG} in the following way.  In orbital space, a perturbation may be diagonal (such that it does not mix the $d_{xz}$ and the $d_{yz}$ orbits) or off diagonal.  We follow Ref.\onlinecite{ZhangQPI} and focus on orbitally diagonal perturbation.  Our conclusion does not depend on this choice.  In the Nambu space diagonal disorder represents impurities that may couple to charge (non-magnetic impurity, $\tau_3$) or spin (magnetic impurity, $\tau_0$).  These two perturbations are closely related and we present here only the non-magnetic case.  Off diagonal perturbations in Nambu space are related to  pairing, we shall see below that the presence of a vortex takes this form.

A point-like non-magnetic impurity is simply described by:
\begin{equation}
\delta{\cal H} = V_0\sum_{\sigma}(c^\dagger_{\sigma}(\br_0)c_{\sigma}(\br_0)+d^\dagger_{\sigma}(\br_0)d_{\sigma}(\br_0)),
\end{equation}
where $c_\sigma(\br)$ represent the electron annihilation operators on site $\br$.
In the language of the two-band Nambu space this leads to  $V_{\bk, \bk'}= V_0\sigma_0\otimes\tau_3$.  This perturbation has been studied in Ref.\onlinecite{ZhangQPI} and we do not wish to repeat the analysis.  Instead we would like to draw the reader's attention to four scattering processes along the x-axis.  They are denoted by their momentum transfer vectors $\bq_1,\bq_2,\bq_3$ and $\bq_4$ in Fig.~\ref{fig:CCElabeled}.  Of these transitions $\bq_1$ and $\bq_2$ are inter-pocket transitions and $\bq_3,\bq_4$ are intra-pocket.  We can clearly see these transitions in the dashed (online red) curves in Fig~\ref{fig:cuts}, in both panels.  Moreover, one may notice that in the presence of a non-magnetic impurity intra- and inter-pocket transitions have opposite signs of intensity in the $s_\pm$ scenario and the same sign in the simple $s$-wave case.  This could have, in principle, served as a way to distinguish between the two OPs.  However, this may be problematic because
(i) Scanning tunneling spectroscopy experiments are {\it not} sensitive to this sign\cite{Capriotti} and (ii) The magnetic scattering channel ($\tau_0$) mixes in with the non-magnetic ($\tau_3$) even though the impurity has no magnetic moment when scattering is strong\cite{Balatsky}.  We shall show below that the analysis that makes use of the vortex induced scattering is free of these problems.

Let us consider a magnetically induced vortex.  We simplify its effect by considering only the {\em amplitude} suppression of the order parameter near the vortex core.  This simplification is crucial for our analysis; including the OP phase winding leads to technical complications which, at the moment, we are unable to overcome.  Nevertheless, the same simplification has been carried out for a $d$-wave superconductor and proved useful in explaining experimental results\cite{TPBMFfield}.  We consider a 'point-like vortex' by suppressing the OP amplitude only on four next-nearest neighbor links around a single site (labeled zero).  Larger cores may be taken into account by adding up a few of these perturbations.  The modification to the Hamiltonian caused by the vortex is described by:
\begin{eqnarray}
\delta {\cal H} &=& \sum_\delta \delta \Delta [ c_{\uparrow}(r_0) c_\downarrow(r_0 + \hat{\delta}) - c_\downarrow(r_0) c_\uparrow(r_0 + \hat{\delta}) \nonumber \\
 &+& (c  \rightarrow  d)+  h.c. ],
\end{eqnarray}
where $\hat{\delta} = \pm \hat{x} \pm \hat{y}$. In the two-band-Nambu basis we get:
\begin{equation}
V_{\bk \bk'} = 4 \delta \Delta \left( \chi_\bk + \chi_{\bk'} \right)\sigma_0\otimes\tau_1
\end{equation}
where $\chi_\bk = \cos k_x \cos k_y = \Delta_\pm(\bk)/\Delta_0$ is the result of next nearest neighbor summation.

%
\begin{widetext}
\
\begin{figure}
\includegraphics[width=0.45\columnwidth]{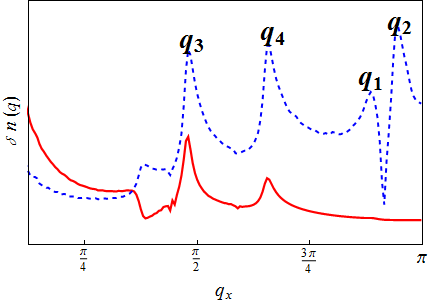}
\hfill
\includegraphics[width=0.45\columnwidth]{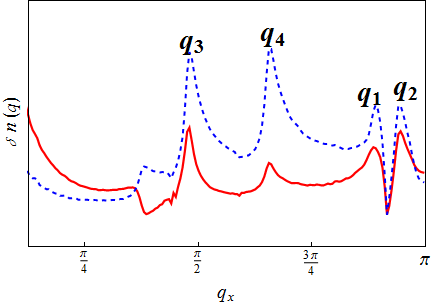}
\caption{{\bf Quasiparticle interference cuts. }
The tunneling density of states modulations $\delta n(\bq)$ plotted along $q_y=0$ for $q_x \in \left(0, \pi \right)$. The transitions $\bq_i$ are the same as in Fig.~\ref{fig:CCElabeled}. Both plots are for the same parameters as in Fig.~\ref{fig:CCElabeled}: $\mu=1.5, \omega = 0.105$.
In both pallets the solid (red) line represents vortex scattering and the dashed (blue) line is for impurity scattering. Both curves of pallet (a) were generated with the $s_\pm$ order parameter while, for comparison, the curves in pallet (b) was generated using the absolute value $|\Delta_\pm(\bk)|$. The presence of peaks at $\bq_1$ and $\bq_2$ in these plots is a confirmation that the sign change of $s_\pm$ play a crucial role in their suppression in pallet (a).
}
\label{fig:cuts}
\end{figure}
\end{widetext}

\section{Feature Intensities}
\subsection{Intra- and Inter-pocket scattering}
Next we turn to evaluate the induced modulations in the LDOS in both cases of scattering.  In order to distinguish between intra- and inter-pocket scattering processes it is useful to work in the pocket (energy band) basis rather than the orbital basis.  To do this we transform the $4\times4$ matrices of the Green's functions and potentials through the unitary transformation:
\begin{eqnarray}
U(\bk) = \left(  \cos(\beta_\bk / 2)\sigma_0 - i \sin(\beta_\bk / 2) \sigma_2 \right)\otimes\tau_0 \nonumber \\
\beta_{\bk} = \arctan\left( {\epsilon_{xy}(\bk) \over \epsilon_-(\bk)} \right).
\end{eqnarray}
This transformation diagonalizes the kinetic part of the Hamiltonian: $\hat{h}_t$. We define the Green's function and the potential in the pocket basis by:
\begin{eqnarray}
\tilde G^0(\bk,\omega) = U(\bk)^{-1} G^0(\bk,\omega)U(\bk) \nonumber \\
\tilde V_{\bk,\bk'} = U(\bk)^{-1} V_{\bk,\bk'}U(\bk')
\end{eqnarray}
Applying the above transformations to $\delta G(\bk,\bk',\omega)$ in the Born approximation (Eq.~\ref{eq:GVG}) we find:
\begin{widetext}
\begin{equation}\label{eq:deltaG}
\delta G(\bk, \bk+\bq, \omega)|_{11+33} = V_0\left(\left[ P_1(\bk,\bq) + P_2(\bk, \bq) \right]\cos^2 \frac{\beta_\bk - \beta_{\bk+\bq}}{2} + \left[ Q_1(\bk,\bq) + Q_2(\bk, \bq) \right] \sin^2 \frac{\beta_\bk - \beta_{\bk+\bq}}{2}\right).
\end{equation}
\end{widetext}
In the above expression the sine and cosine functions are the result of the transformation.
The 'angle' represents the amount of orbital mixing in the bands.  It is interesting to note that on the $\hat x$ and $\hat y$ axes the orbitals are not mixed (since $\epsilon_{xy}=0$).  This prohibits inter-band on-shell transitions on the axis.  However, intensity around the vectors $\bq_1$ and $\bq_2$ is still non-vanishing due to the contribution of close-by off-shell states.
The functions $P_{1(2)}(\bk,\bq)$ represent intra-pocket transitions within the kinetic energy band $\epsilon_{1/2}(\bk) = \epsilon_+(\bk) \pm \sqrt{ \epsilon^2_-(\bk)+\epsilon^2_{xy}(\bk)}$ and the function $Q_{1(2)}(\bk,\bq)$ represent transitions from one band to another.  In the case of a non-magnetic impurity we find:
\begin{widetext}
\begin{eqnarray}
P_{1(2)} &=& \frac{ (\omega + \epsilon_{1(2)}(\bk))(\omega + \epsilon_{1(2)}(\bk+\bq)) - \Delta(\bk) \Delta(\bk + \bq)}{(\omega^2 - \epsilon_{1(2)}(\bk)^2 - \Delta(\bk)^2+i\eta)(\omega^2 - \epsilon_{1(2)}(\bk+\bq)^2 - \Delta(\bk+\bq)^2+i\eta)}
\\
Q_{1(2)} &=&  \frac{ (\omega + \epsilon_{1(2)}(\bk))(\omega + \epsilon_{2(1)}(\bk+\bq)) - \Delta(\bk) \Delta(\bk + \bq)}{(\omega^2 - \epsilon_{1(2)}(\bk)^2 - \Delta(\bk)^2+i\eta)(\omega^2 - \epsilon_{2(1)}(\bk+\bq)^2 - \Delta(\bk+\bq)^2 +i\eta)}
\end{eqnarray}
\end{widetext}
By tuning the energy (or the external bias in the experiment) to the gap edge we may assume that the kinetic energy is roughly equal to the chemical potential such that $\epsilon_i(k) \sim 0$.  This simplify the functions $P$ and $Q$ to:
\begin{eqnarray} \label{qp-nm}
P_i(\bk,\bq) &\approx& Q_i(\bk, \bq)   \\ \nonumber
&\approx&\frac{\omega^2 - \Delta(\bk) \Delta(\bk+\bq)}{(\omega^2 - \Delta(\bk)^2+i\eta)(\omega^2 - \Delta(\bk+\bq)^2+i\eta)}
\end{eqnarray}
Note that the imaginary part of the above expression (which contributes to the observed LDOS) is an odd function of $\omega$.

A similar derivation can be done for the vortex perturbation.  Eq.~\ref{eq:deltaG} remains the same except for the replacement of $V_0$ by $\delta\Delta$ and the functions $P$ and $Q$ are:
\begin{widetext}
\begin{eqnarray}
P_{1(2)} = \frac{ \Delta(\bk)(\omega + \epsilon_{1(2)}(\bk+\bq)) + \Delta(\bk+\bq)(\omega + \epsilon_{1(2)}(\bk))}{(\omega^2 - \epsilon_{1(2)}(\bk)^2 - \Delta(\bk)^2+i\eta)(\omega^2 - \epsilon_{1(2)}(\bk+\bq)^2 - \Delta(\bk+\bq)^2+i\eta)}(\chi_\bk + \chi_{\bk+\bq})
\\
Q_{1(2)} =  \frac{ \Delta(\bk)(\omega + \epsilon_{1(2)}(\bk+\bq)) + \Delta(\bk+\bq)(\omega + \epsilon_{2(1)}(\bk))}{(\omega^2 - \epsilon_{1(2)}(\bk)^2 - \Delta(\bk)^2+i\eta)(\omega^2 - \epsilon_{2(1)}(\bk+\bq)^2 - \Delta(\bk+\bq)^2+i\eta)}(\chi_\bk + \chi_{\bk+\bq})
\end{eqnarray}
\end{widetext}
Where the differences arise from the different matrix structure of the potential ($\tau_1$ in this case as opposed to $\tau_3$ in the previous case).
Again, when we tune the energy to obtain $\epsilon_i(k) \sim 0$ the expressions simplify to:
\begin{eqnarray} \label{qp-vortex}
P_i(\bk,\bq) &\approx& Q_i(\bk, \bq) \\ \nonumber
&\approx& \frac{\omega (\Delta(\bk) + \Delta(\bk+\bq))^2/ \Delta_0 }{(\omega^2 - \Delta(\bk)^2+i\eta)(\omega^2 - \Delta(\bk+\bq)^2+i\eta)}
\end{eqnarray}
In this case the observed quantity will be an even function of the bias voltage, $\omega$ and this in principle can be used to distinguish between the two sources of scattering.
In Eq.~\ref{eq:deltan} we integrate over the Brillouin zone such that $\bk$ and $\bk+\bq$ go over all pairs of states that are separated by momentum $\bq$.  The main contribution to this sum comes from the vicinity of the contours of constant energy such as the beginning and end points of the arrows in Fig.~\ref{fig:CCElabeled}.  This means that the intensity of the FT LDOS features in momentum $\bq$ depend on the gap function at these two points.  It is obvious from Eq.~\ref{qp-vortex} that when $\bq$ connects two points on the {\em same pocket} the vortex induced scattering will be large and will be added to the impurity scattering (it may enhance or suppress it depending on the relative signs of $V_0$ and $\delta\Delta$ and the sign of the bias voltage).  On the other hand, if $\bq$ connects two points on {\em different} pockets the contribution will vanish if $\Delta(\bk_i) \approx -\Delta(\bk_f)$ as expected in the $s_\pm$ scenario.

This is the main finding of this paper - if the OP of the FeSC is of the $s_{\pm}$ structure and has roughly the same magnitude on both the electron and hole pockets the application of magnetic field will affect only features at momenta $\bq$ which correspond to intra-pocket transitions.  Inter-pocket transitions will stay intact.  To best see this effect the bias voltage should be tuned to $\pm\Delta_0$ and the odd and even components of the LDOS should be separated.

\subsection{Full lattice model results}
\begin{figure}
\includegraphics[width=0.48\columnwidth]{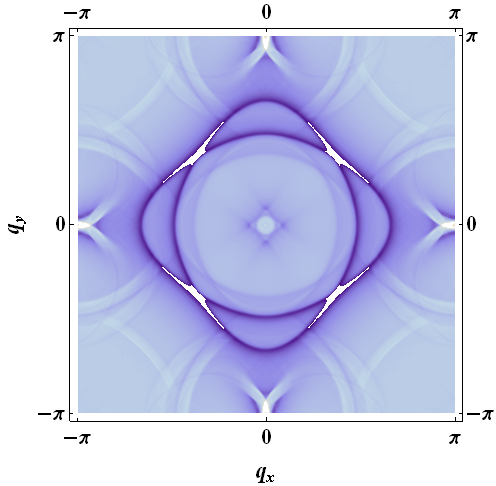} \hfill
\includegraphics[width=0.48\columnwidth]{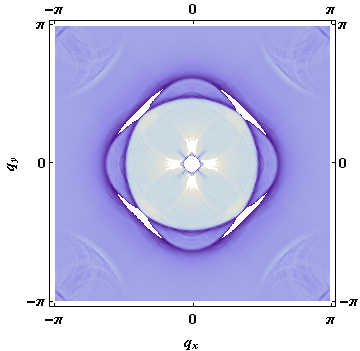}
\caption{{\bf Quasiparticle interference patterns in the Brillouin zone. }
(Color online) $\delta n(\bq)$ for $\mu=1.5$ and $\omega = 0.105$ on a 400 x 400 lattice.  Left: Non-magnetic impurity induced interference patterns, Right: Vortex induced interference patterns.  The patterns are similar except for the features close to $(\pm\pi,0)$ and $(0,\pm\pi)$ where the inter-pocket transitions reside.  It is clear that these transitions are missing from the modulations generated by the vortex.
}
\label{QPI-figure}
\end{figure}

In order to account for both on-shell and off-shell states in an exact fashion, we performed a numerical study of Eq.~\ref{eq:deltaG}.  A sample of our results in the Brillouin zone is presented as a grey scale plot in Fig.~\ref{QPI-figure} and in cuts along the $\hat x$ axis in Fig.~\ref{fig:cuts}.  We identify the transitions $\bq_i$ as labeled in Fig.~\ref{fig:CCElabeled} by their momentum transfer.  In order to ascertain that the transitions are correctly identified we vary parameters (like the band parameters, the chemical potential and the energy) and follow the transitions' evolution in momentum space.

In Fig.~\ref{fig:cuts}a,  it is clear that while sharp peaks appear in the LDOS at the inter-pocket momenta $\bq_1$ and $\bq_2$ in the case of a non-magnetic impurity, they are absent in the case of a vortex scatterer. For comparison, when replacing the $s_\pm$ OP by a function without a sign change (absolute value) the inter-pocket transitions appear in both types of scatter.

\section{Discussion}
We have shown that an order parameter which changes sign between the hole and electron pockets of the FeSCs (such as the proposed $s_\pm$) could be distinguished from an order parameter which does not change sign (a simple or anisotropic $s$-wave).  We propose to measure the modulations in the LDOS due to scattering off impurities and magnetic vortices.  The conclusion of our analysis is that the features of the vortex-induced scattering are sensitive to the OP magnitude and sign. In particular, when the energy is tuned to the gap edge, the intensity of the LDOS features at momentum $\bq$ are roughly proportional to $(\Delta(\bk_i)+\Delta(\bk_f))^2$ where $\bk_i$ and $\bk_f$ are two points on the relevant contours of constant energy which are separated by momentum $\bq$.  This means that if the OP is of similar magnitude but opposite sign on the two points a cancelation will occur.  Such cancelation is expected for the $s_\pm$ OP but not for an $s$-wave, and is therefore a signature of the $s_\pm$ OP.

It is important to emphasize that the conclusion of the analysis presented in this work is
quite robust. Numerical simulations with modified lattice parameters and chemical potential has yielded modified
quasiparticle interference patterns. However, the absence of the transitions $\bq_1$ and $\bq_2$ appears in the vortex scattering
contribution whenever the OP changes sign between the electron and hole pockets.
We would like to mention that the Born approximation, chosen for the current study, is both simple and
appropriate when the scattering potential is weak. For stronger scattering the $t$-matrix may be of use. It's
strength is generally in finding impurity bound state\cite{BoundStates}
or vortex bound states (Caroli-de-Gennes-Matricon)\cite{NagaiKato}. In general, it amounts to replacing the bare scattering
matrix by an energy dependent one. Often, the $t$-matrix
does not add any momentum dependence and therefore
will lead to the same quasiparticle interference patterns
as the Born approximation.

The proposed experiment is scanning tunneling spectroscopy in magnetic field.  Such measurements have been successfully carried out in the cuprates\cite{Hanaguri} and have been interpreted in a similar fashion to that proposed here\cite{TPBMFfield}.  We are hopeful that advances in material synthesis will allow similar studies of the FeSCs.  Their quasi-two dimensional structure is ideal for STM and their transition temperature is high enough for such experiments.  An important consideration is the vortex core size in FeSC.  In general, the larger the core size, the harder it is to detect features at large momenta (close to the BZ edges).  In the Born approximation, in order to replace the point-like vortex considered here by a more realistic one, the perturbation $\delta\Delta$ should be replaced by a more smoothly varying function in real space with a characteristic length scale $\xi$, the coherence length.  Its Fourier transform $\delta\Delta(\bq)$ will multiply our result for $\delta n(\bq,\omega)$ creating an envelope beyond which features are suppressed.  In BaFe$_{1.8}$Co$_{0.2}$As$_2$ the coherence length was measured to be $\sim 27.6 \AA$\cite{HoffmanPnictides}, which may lead to a significant signal suppression at $(\pi,0)$.  However, this coherence length is similar to that of BiSCCO (a member of the cuprate family)\cite{Pan} where FT LDOS features were seen clearly even at the edges of the Brillouin zone\cite{Hoffman}.  This suggests that the vortex size and structure in BaFe$_{1.8}$Co$_{0.2}$As$_2$ may be suitable for the suggested experiment.

\section{Acknowledgements}
The authors would like to acknowledge useful discussions with M. Franz, O. Motrunich and Z. Te\v{s}anovi\'{c}, the hospitality of the Aspen center for physics were some of the work has taken place and funding from the Caltech SURF
program (EP); the Packard and Sloan Foundations, the Institute for Quantum Information under NSF grants PHY-0456720 and PHY-0803371, and The Research Corporation Cottrell Scholars program (GR).
\bibliographystyle{apsrev}
\bibliography{final}
\end{document}